%% file: main.tex
\documentclass[12pt]{article}
\usepackage{fullpage}

\usepackage[sectionbib,authoryear]{natbib}
\usepackage{graphicx}
\usepackage{amsmath}
\usepackage{url}

\begin{document}


\title{DeepNano: Deep Recurrent Neural Networks for Base Calling in MinION Nanopore Reads}
\author{Vladim\'\i{}r Bo\v{z}a,
Bro\v{n}a Brejov\'a and Tom\'a\v{s} Vina\v{r}}
\date{Faculty of Mathematics, Physics and Informatics,
           Comenius University,\\
           Mlynsk\'a dolina, 842 48 Bratislava, Slovakia
}




\maketitle

\begin{abstract}
\textbf{Motivation:} The MinION device by Oxford Nanopore is the first
portable sequencing device. MinION is able to produce very long reads (reads
over 100~kBp were reported), however it suffers from high sequencing error
rate. In this paper, we show that the error rate can be reduced by improving
the base calling process.\\
\textbf{Results:} We present the first open-source DNA base caller for
the MinION sequencing platform by Oxford Nanopore.  By employing
carefully crafted recurrent neural networks, our tool improves the
base calling accuracy compared to the default base caller supplied by
the manufacturer.  This advance may further enhance applicability of
MinION for genome sequencing and various clinical applications.\\
\textbf{Availability:} DeepNano can be downloaded at 
\url{http://compbio.fmph.uniba.sk/deepnano/}.\\
\textbf{Contact:} boza@fmph.uniba.sk
\end{abstract}

\input intro.tex

\input model.tex
\input exp.tex

\input concl.tex

\paragraph{Acknowledgements.} This research was funded by
VEGA grants 1/0684/16 (BB) and 1/0719/14 (TV), and a grant
from the Slovak Research and Development Agency APVV-14-0253.
The authors thank Jozef Nosek for involving us in the MinION Access Programme,
which inspired this work.

\bibliographystyle{apalike} \bibliography{main}

\end{document}

%% file: intro.tex
\section{Introduction}

In this paper, we introduce the first open-source base caller for the
MinION nanopore sequencing platform \citep{minion1}. 
The MinION device by Oxford Nanopore, weighing only 90 grams, is
currently the smallest high-throughput DNA sequencer. 
Thanks to its low capital costs, small size and the possibility 
of analyzing the data in real time as they are produced, MinION is very
promising for clinical applications, such as monitoring infectious
disease outbreaks \citep{Judge2015,Quick2015,Quick2016},
and characterizing structural variants in cancer \citep{Norris2016}.
Although MinION
is able to produce long reads, they have a high sequencing
error rate.  In this paper, we show that this error rate can be
reduced by replacing the default base caller provided by the
manufacturer with a properly trained deep recurrent neural network.

\begin{figure}
\centering
\includegraphics[width=0.8\columnwidth]{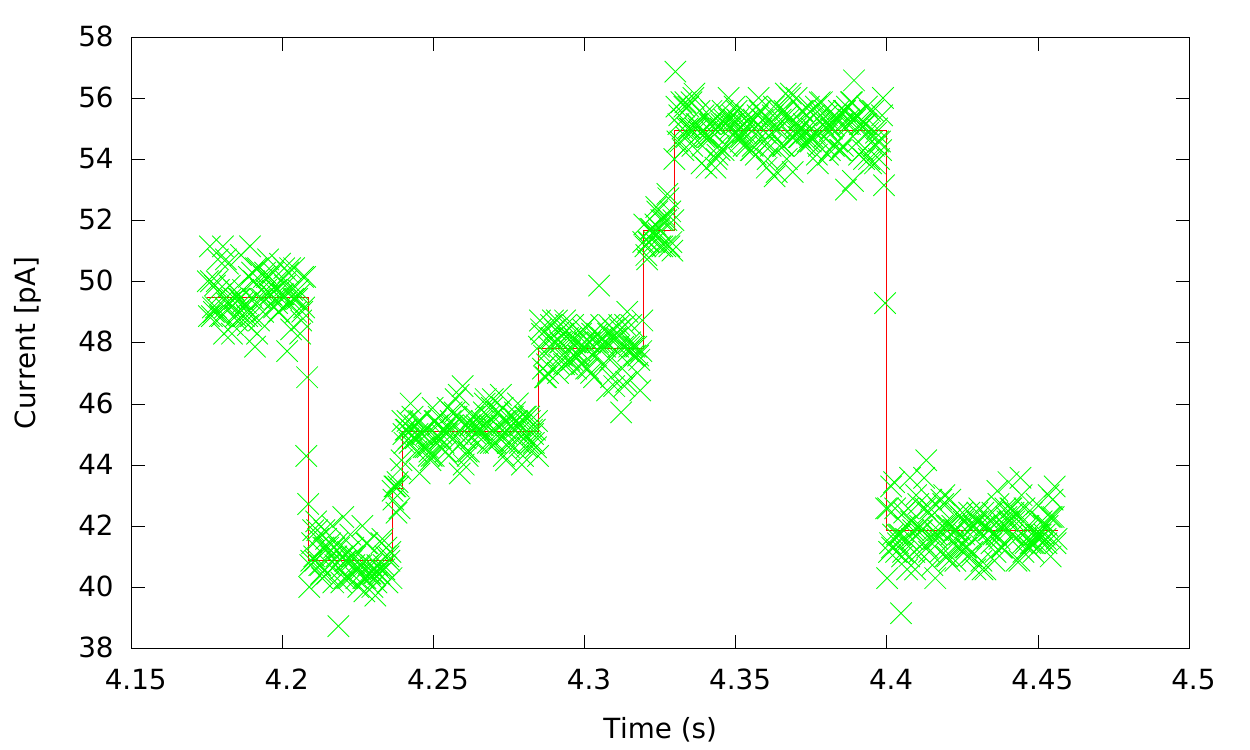}
\caption{{\bf Raw signal from MinION 
and its segmentation to events.}
The plot was generated from the \emph{E. coli} data \citep{ecolidata}.}
\label{fig:rawdata}
\end{figure}

In the MinION device, single-stranded DNA
fragments move through nanopores, which causes drops in the
electric current. 
The electric current is measured at each pore several thousand times
per second, resulting in a measurement plot 
as shown in Fig.\ref{fig:rawdata}. The
electric current depends mostly on the context of several DNA bases passing
through the pore at the time of measurement. As the DNA moves through
the pore, the context shifts and the electric current changes. Based
on these changes, the sequence of measurements is split into
\emph{events}, each event ideally representing the shift of the context
by one base. Each event is summarized by the mean and variance 
of the current and by event duration. 
This sequence of events is then translated
into a DNA sequence by a base caller.

A MinION device typically yields reads several thousand bases long;
reads as long as 100,000~bp have been reported.  To reduce the error
rate, the device attempts to read both strands of the same DNA
fragment. The resulting template and complement reads can be combined
to a single two-directional (2D) 
read during base calling. As shown in Table \ref{table:exp},
this can reduce the error rate of the default base caller from roughly 30\%
for 1D reads to 13-15\% for 2D reads. 

The default base caller provided by Oxford Nanopore is called Metrichor.
It is a proprietary software, and the exact details of
its algorithms are not known. It assumes that each event depends on 
a context of $k=6$ consecutive bases 
and that the context typically shifts by one base in
each step. As a result, every base is read as a part of $k$
consecutive events.

This process can be represented by a hidden Markov model (HMM). Each state in
the model represents one $k$-tuple and the transitions between
states correspond to $k$-tuples overlapping by $k-1$ bases (e.g.
AACTGT will be connected to ACTGTA, ACTGTC, ACTGTG, and ACTGTT),
similarly as in de~Bruijn graphs. Emission probabilities
reflect the current expected for a particular $k$-tuple, with an
appropriate variance added. Finally, additional transitions need to be
added, representing missed events, falsely split events, and other likely
errors (in fact, insertion and deletion errors are quite common in the
MinION sequencing reads, perhaps due to errors in event segmentation).
After parameter training, base calling can be performed by running the
Viterbi algorithm, which will result in the sequence of states with
the highest likelihood. It is not known, what is the exact nature of
the model used in Metrichor, but the emission probabilities required
for this type of model are provided by Oxford Nanopore in the files
storing the reads.

In our custom base caller DeepNano, 
we opt to use recurrent neural networks, which
have stellar results for speech recognition \citep{rnnspeech}, machine translation
\citep{rnntranslation}, language modeling \citep{rnnlang}, and other sequence
processing tasks. Note that neural networks were previously used
for base calling Sanger sequencing reads \citep{Tibbetts1994,Mohammed2013},
though the nature of MinION data is rather different. 

Several tools for processing nanopore sequencing data were already
published, including read mappers \citep{Jain2015,Sovic2015}, and
error correction tools using short Illumina reads \citep{Goodwin2015}.  Most
closes related to our work are Nanopolish \citep{loman2015} and PoreSeq
\citep{Szalay2015}. Both of these tools create a consensus sequence by
combining information from multiple overlapping reads, considering not
only the final base calls from Metrichor, but also the sequence of events.
They analyze the events by hidden Markov models with emission
probabilities provided by Metrichor. In contrast, our base caller does
not require read overlaps, it processes reads individually and provides
more precise base calls for downstream analysis. The crucial
difference, however, is our use of a more powerful discrimitative
framework of recurrent neural networks.  Thanks to a large hidden
state space, our network can potentially capture long-distance
dependencies in the data, whereas HMMs use fixed $k$-mers.

%% file: model.tex
\section{Basecalling using deep recurrent neural networks}
\label{sec:overview}

In this section, we describe the design of our basecaller, which is
based on deep recurrent neural networks. 
A recurrent neural network \citep{rnn, gravesrnn} is a type of artificial
neural network used for sequence labeling. Given
a sequence of input vectors $\{\vec{x}_1, \vec{x}_2, \dots,
\vec{x}_t\}$, its prediction is a sequence of output 
vectors $\{\vec{y}_1, \vec{y}_2, \dots, \vec{y}_t\}$.
In our case, the inputs vectors consist of the mean, standard 
deviation and length of each
event, and the output vectors give a probability distribution of called bases.

\subsection{Simple recurrent neural networks}

First, we describe a simple neural network with one hidden layer.
During processing of each input vector $\vec{x}_i$, a recurrent 
neural network calculates two vectors: its hidden state $\vec{h}_i$
and the output vector $\vec{y}_i$. Both depend on 
the current input vector and the previous hidden state:
$\vec{h}_i = f(\vec{h}_{i-1}, \vec{x}_i)$, $\vec{y}_i = g(\vec{h}_i)$. 
We will describe our choice of functions $f$ and $g$ later.
The initial state $\vec{h}_0$ is one of the parameters of the model.

Prediction accuracy can be usually improved by using neural networks
with several hidden layers, where each layer uses hidden states from
the previous layer. We use networks with three or four
layers. Calculation for three layers proceeds as follows:
\begin{eqnarray*}
\vec{h}_i^{(1)} &=& f_1(\vec{h}_{i-1}^{(1)}, \vec{x}_i)\\
\vec{h}_i^{(2)} &=& f_2(\vec{h}_{i-1}^{(2)}, \vec{h}_i^{(1)})\\
\vec{h}_i^{(3)} &=& f_3(\vec{h}_{i-1}^{(3)}, \vec{h}_i^{(2)})\\
\vec{y}_i &=& g(\vec{h}_i^{(3)})
\end{eqnarray*}

Note that in different layers, we use different functions $f_1$, $f_2$,
and $f_3$, where each function has its own set of parameters.

\subsection{Bidirectional recurrent neural networks}

\begin{figure}
\centering
\includegraphics[width=0.4\columnwidth]{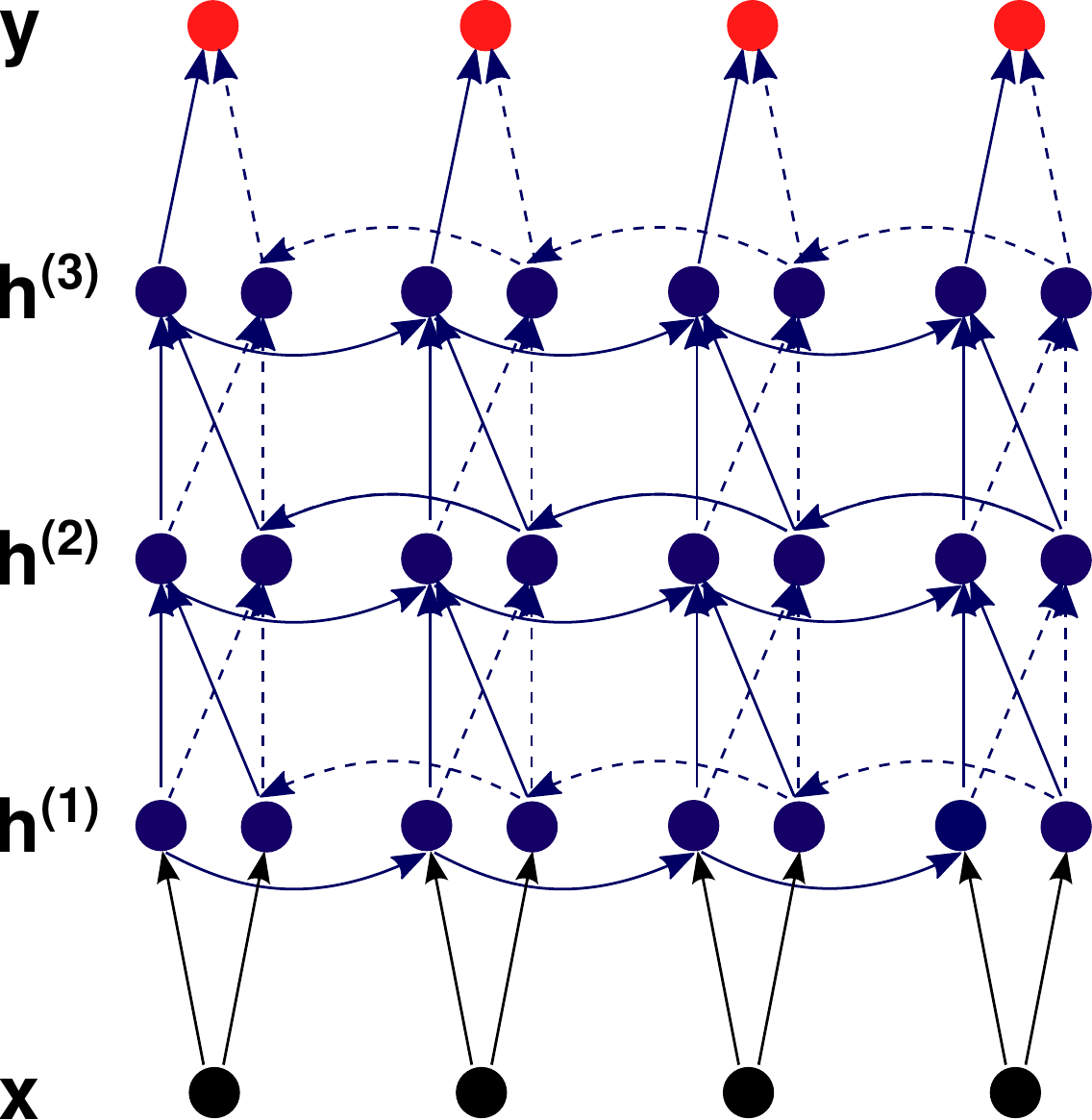}
\caption{Schematics of a bidirectional recurrent neural network}
\label{fig:brnn}
\end{figure}

In our case, the prediction for input vector $\vec{x}_i$
can be influenced by data seen before $\vec{x}_i$ but also by data 
seen after it. To incorporate this data into prediction, we use a
bidirectional neural network \citep{brnn}, which scans data in both
directions and concatenates hidden outputs before proceeding to the
next layer (see Fig. \ref{fig:brnn}). Thus, for a two-layer network,
the calculation would proceed as follows 
($||$ denotes concatenation of vectors):
\begin{eqnarray*}
\vec{h}_i^{(1+)} &=& f_{1+}(\vec{h}_{i-1}^{(1+)}, \vec{x}_i)\\
\vec{h}_i^{(1-)} &=& f_{1-}(\vec{h}_{i+1}^{(1-)}, \vec{x}_i)\\
\vec{h}_i^{(1)} &=&  \vec{h}_i^{(1+)} || \vec{h}_i^{(1-)}\\
\vec{h}_i^{(2+)} &=& f_{2+}(\vec{h}_{i-1}^{(2+)}, \vec{h}_i^{(1)})\\
\vec{h}_i^{(2-)} &=& f_{2-}(\vec{h}_{i+1}^{(2-)}, \vec{h}_i^{(1)})\\
\vec{h}_i^{(2)} &=&  \vec{h}_i^{(2+)} || \vec{h}_i^{(2-)}\\
\vec{y}_i &=& g(\vec{h}_i^{(2)})\\
\end{eqnarray*}

\subsection{Gated recurrent units}

The typical choice of function $f$ in a recurrent neural network
is a linear transformation of inputs followed by hyperbolic tangent nonlinearity:
$$f(\vec{h_{i-1}}, \vec{x_i}) = \tanh(W \vec{x_i} + U \vec{h}_{i-1} + \vec{b})$$
where the matrices $W, U$ and the bias vector $\vec{b}$ are parameters
of the model. Note that we use separate parameters for each layer and
direction of the network.

This choice unfortunately leads to the vanishing gradient problem \citep{van}.
During parameter training, the gradient of the error function in
layers further from the output is much smaller that in layers
closer to the output. In other words, gradient diminishes during
backpropagation through network, 
complicating training of the network.

One of the solutions is to use gated recurrent units in
the network \citep{gru}. Given input $\vec{x_i}$ and previous hidden state
$\vec{h}_{i-1}$, a gated recurrent unit first calculate values for
update and reset gates:
$$\vec{u}_i = \sigma(W_u \vec{x_i} + U_u \vec{h}_{i-1} + b_u),$$
$$\vec{r}_i = \sigma(W_r \vec{x_i} + U_r \vec{h}_{i-1} + b_r),$$
where $\sigma$ is the sigmoid function: $\sigma(z) = 1/(1 + e^{-z})$.
Then the unit computes a potential new value 
$$\vec{n}_i = \tanh(W \vec{x_i} + \vec{r_i} \circ U \vec{h}_{i-1}).$$
Here, $\circ$ is the element-wise vector product. If some component of
the reset gate vector is close to 0, it decreases the impact of the
previous state. 

Finally, the overall output is a linear combination of $\vec{n}_i$ and
$\vec{h}_{i-1}$, weighted by the update gate vector $\vec{u_i}$:
$$\vec{h}_i = \vec{u_i} \circ \vec{h}_{i-1} + (1 - \vec{u}_i) \circ \vec{n}_i.$$

Note that both gates give values from interval $(0, 1)$ and allow
for a better flow
of the gradient through the network, making training easier. 

\subsection{Output layer}

Typically, one input event leads to one called base.
But sometimes we get multiple
events for one base, so there is no output for some events. 
Conversely, 
some events are lost, and we need to call multiple bases for one event.
We limit the latter case to two bases per event.
For each event, we output two probability distributions over the alphabet 
$\Sigma = \{A,C,G,T,-\}$, where the dash means no base. 
We will denote the two bases predicted for input event $\vec{x}_i$ as 
$b_i^{(1)}$ and $b_i^{(2)}$.
Probability of each base $q\in\Sigma$ 
is calculated from the hidden states in the last layer 
using the softmax function:
$$P[b_i^{(k)} = q] = \frac{\exp(\vec{\theta}_q^{(k)} \vec{h}_{i}^{(3)})}{
\sum_{p \in \Sigma} \exp(\vec{\theta}_p^{(k)} \vec{h}_{i}^{(3)})}$$
Final basecalling is done by taking the most probable base 
for each $b_i^{(k)}$ (or no base if dash is the most probable
character from $\Sigma$). 
During training, if there is one base per event, 
we always set $b_i^{(1)}$ to dash.

\subsection{Training}

Let us first consider the scenario in which we know the correct DNA
bases for each input event.  
The goal of the training is then to find parameters of the network that
maximize the log likelihood of the correct outputs. In particular, if 
$o_1^{(1)}, o_1^{(2)}, o_2^{(1)}, \dots, o_n^{(1)}, o_n^{(2)}$
is the correct sequence of output bases, we try to maximize the sum
$$\sum_{i=1}^n \lg P[b_i^{(1)} = o_1^{(1)}] + \lg P[b_i^{(2)} = o_2^{(2)}]$$

As an optimization algorithm, we use stochastic gradient descent (SGD)
combined with Nesterov momentum \citep{mom} to increase the convergence rate.
For 2D basecalling, we first use SGD with Nesterov momentum,
and after several iterations we switch to L-BFGS \citep{lbfgs}.
Our experience suggests that SGD is better at avoiding bad local 
optima in the initial phases of training, while
L-BFGS seems to be faster during the final fine-tuning.

Unfortunately, we do not know the correct output sequence; 
more specifically, we only know the region of the reference sequence
where the read is aligned, but we do not know the exact pairs
of output bases for individual events. 
We solve this problem in an EM-like fashion.
First, we create an approximate alignment between the events and the
reference sequence using a simple heuristic (we try to minimize the difference
between the expected and observed means for events and have simple penalties for undetected and split
events).
Then every hundredth epoch of optimization, we realign the events to
the reference sequence.
We score the alignment by computing the log-likelihood of bases
aligned to each event in the probability distribution produced by the
current version of the network.
To find the alignment with maximum likelihood, we use a 
simple dynamic programming.

\subsection{1D basecalling}

The neural networks described above can be used for basecalling template and complement strands
in a straightforward way. 
Note that we need a separate model for each strand, since they
have different properties.
In both models, we use neural networks with three hidden layers and 100 hidden units.

As an input for the networks, we use event data stored in Fast5 files 
produced by Metrichor,
and apply the preprocessing parameters for scaling and shifting the signal from the events, also stored in Fast5 files. 

\subsection{2D basecalling}

In 2D basecalling, we need to combine information from separate event sequences for the template and complement strands. A simple option is to apply neural networks for each strand separately, producing two sequences of output probability distributions. Then we can align these two sequences of distributions by dynamic programming and produce the DNA sequence with maximum likelihood. 

However, this approach leads to unsatisfactory results in our models, with the same or slightly worse accuracy than the original Metrichor basecaller. We believe that this phenomenon occurs because our models output independent probabilities
for each base, while the Metrichor basecaller allows dependencies between adjacent basecalls.

Therefore, we have built a neural network which gets as an input
corresponding events from the two strands and combines them to a
single prediction. To do so, we need an alignment of the two event sequences,
as some events can be falsely split or missing in one of the
strands. We use the alignment obtained from the base call files produced by
Metrichor. We convert each pair of aligned events to a single
input vector. Events present in only one strand are completed to a
full input vector by special values.
This input sequence is then used in a neural network with four hidden
layers and 250 hidden units in each layer.

\subsection{Implementation details}

We have implemented our network using Theano library for Python \citep{theano}, which includes symbolic differentiation
and other useful features for training neural networks. We do not use any regularization, as with the size of our dataset
we saw almost no overfitting. 

%% file: exp.tex
\section{Experimental results}

\subsection{Data sets}

We have used two existing sets of bacterial reads produced by the SQL-MAP006 sequencing protocol.
The first set of reads is from the genome of \emph{Escherichia coli} 
\citep{ecolidata} and the second one from \emph{Klebsiella pneumoniae} 
\citep{klebsielladata}.
We only used reads which passed the original base calling process and
had a full 2D base call. We have also omitted reads that did not map
to the reference sequence (mapping was done separately for 2D
base calls and separately for base calls from individual strands).

We have split the \emph{E.~coli} data set into training and testing portions; the training set contains the reads mapping to the first $2.5$ Mbp of the genome.
We have tested the predictors on reads which mapped to the rest of the \emph{E.~coli} genome and on reads from \emph{K. pneumoniae}.
Basic statistics of the two data sets are shown in Table \ref{table:sum}.

\begin{table}
\caption{{\bf Sizes of experimental data sets.} The sizes differ between strands because only base calls mapping to the reference were used. Note that the counts of 2D events are based on the size of the alignment.}
\centering
\begin{tabular}{l@{\qquad}r@{\qquad}r@{\qquad}r}
\hline
                    &   \emph{E.~coli} & \emph{E.~coli}  & \emph{K.~pneumoniae}\\

                       & training & testing & testing \\\hline
\# of template reads    & 3,803           & 3,942          & 13,631      \\
\# of template events   & 26,403,434       & 26,860,314      & 70,827,021   \\
\# of complement reads  & 3,820           & 3,507          & 13,734      \\
\# of complement events & 24,047,571       & 23,202,959      & 67,330,241   \\
\# of 2D reads          & 10,278          & 9,292          & 14,550      \\
\# of 2D events         & 84,070,837  & 75,998,235  & 93,571,823\\
\hline
\end{tabular}
\label{table:sum}
\end{table}

\subsection{Results}

We have compared our base calling accuracy with the accuracy of the original Metrichor base caller.
The main experimental results are summarized in Table \ref{table:exp}. We see that in the 1D case,
our base caller is significantly better on both strands and in both data sets. In 2D base calling,
our accuracy is slightly higher.

\begin{table}
\caption{{\bf Accuracy of base callers on two testing data sets.} The results of
base calling were aligned to the reference using BWA-MEM \citep{bwamem}.
The accuracy was computed as the number of matches in the alignment 
divided by the length of the alignment.
}
\centering
\begin{tabular}{l@{\qquad}rr}
\hline
                    &   \emph{E.~coli}     & \emph{K.~pneumoniae}\\
\hline
\multicolumn{3}{l}{\bf Template reads}\\
Metrichor           & $71.3 \%$     & $68.1 \%$          \\
DeepNano            & $77.9 \%$     & $76.3 \%$          \\[1mm]
\multicolumn{3}{l}{\bf Complement reads}\\
Metrichor           & $71.4 \%$     & $69.5 \%$          \\
DeepNano            & $76.4 \%$     & $75.7 \%$          \\[1mm] 
\multicolumn{3}{l}{\bf 2D reads}\\
Metrichor           & $86.8 \%$     & $84.8 \%$          \\
DeepNano            & $88.5 \%$     & $86.7 \%$          \\
\hline
\end{tabular}
\label{table:exp}
\end{table}

On the \emph{Klebsiella pneumoniae} data set, we have observed a difference in 
the GC content bias between the two programs. 
This genome has GC content of $57.5\%$.
DeepNano has underestimated the GC content
on average by 1\%, whereas the Metrichor 
base caller underestimated it by
2\%. 

\subsection{Base calling speed}

It is hard to compare the speed of the Metrichor base caller with our base caller since,
we do not know the exact configuration of hardware running Metrichor
(Metrichor is a cloud-based service).
From the logs, we are able to ascertain that Metrichor spends
approximately $0.01$ seconds per event during 1D base calling.
DeepNano spends $0.0003$ seconds per event on our server, using one CPU thread.
During 2D base call, Metrichor spends $0.02$ seconds per event 
(either template or
complement), while our base callers spends $0.0008$ seconds per event.
To put these numbers into perspective, base calling a read with 4,962 template and 4,344 complement events takes Metrichor 46s for template, 34s for complement, and 190s for 2D data. DeepNano can process the same read in
$1.5$s for template, $1.3$s for complement, and $11.3$ seconds for 2D data.
We believe that unless Metrichor base calling is done on a highly overloaded server, our base caller has a much superior speed.

Although DeepNano is relatively fast in base calling, it requires
extensive computation during training. The 1D networks were trained 
for three weeks on one
CPU (with a small layer size there was little benefit from parallelism).
The 2D network was trained for three weeks on a GPU, followed by two weeks
of training on a 24-CPU server, as L-BFGS performed better using multiple
CPUs.  Note however that once we train the model for a
particular version of MinION chemistry, we can use the same parameters
to base call all data sets produced by the same chemistry, 
as our experiments indicate that the same parameters
work well for different genomes.

%% file: concl.tex
\section{Conclusion and further research}

In this paper, we have presented a new tool for base calling MinION
sequencing data. Our tool provides a more accurate and computationally
efficent alternative to the HMM-based methods used in the Metrichor
base caller by the device manufacturer.

To further improve the accuracy of our tool, we could explore several 
well-established approaches for improving performance of neural networks. 
Perhaps the most obvious option is to increase the network size.
However, 
that would require more training data to prevent overfitting,
and both training and base calling would get slower.

Another typical technique for boosting the accuracy of 
neural networks is using an ensemble of several networks
\citep{rnntranslation}.  Typically, this is done by training several neural
networks with different initialization and order of training samples,
and then avergaing their outputs.
Again, this technique leads to slower base calling.

The last technique, called dark knowledge \citep{darkknow, darkknow2},
trains a smaller neural network using training data generated from 
output probabilities of a larger
network. The training target of the smaller network is to match
output probability distributions of the larger network. 
This leads to improved performance for
the smaller network compared to training it directly on the training data.
This approach would allow fast base calling with a small network,
but the training step would be time-consuming.

Another avenue for improving our base caller is to remove its 
dependence on the results of Metrichor, which would allow 
users to bypass Metrichor base calling step. 
In particular, to preprocess the events, we use 
scaling parameters generated by Metrichor.
We believe that our networks can handle data without scaling, but this
would require more extensive testing with a variety of sequencing
runs.  During 2D basecalling, we use an alignment between the template
and complement sequences generated by the original basecaller.
To eliminate this dependency, we need to explore several alignment options
of template and complement strand, 
and check how they perform with our basecaller.

Finally, we use event data from Metrichor-generated files, 
but perhaps the accuracy can be futher improved by a
different event segmentation or by using 
more features from the raw signal besides signal mean and standard
deviation. We tried several options (signal kurtosis, difference between 
the first and second halves of
the event, etc.), but the results were mixed.